\documentclass[aps,prb,reprint,showpacs,superscriptaddress]{revtex4-1} % by Martin in Nano Letters draft

\usepackage{graphicx}% Include figure files
\usepackage{dcolumn}% Align table columns on decimal point
\usepackage{bm}% bold math
\usepackage{soul} %Cla

\usepackage{booktabs} % Cla
\usepackage{multirow} % Cla
\usepackage{siunitx} % Cla
\usepackage[margin=1in]{geometry} % Cla

\usepackage{xcolor} % Cla
\usepackage{comment} % Cla
\usepackage{siunitx} % Cla

\usepackage{hyperref} % Cla
\hypersetup{colorlinks=true, citecolor=blue, urlcolor=blue, linkcolor=blue} % Cla

\usepackage[english]{babel} % by Martin in NL Draft
\addto\captionsenglish{}

\begin{document}

\preprint{APS/123-QED}

\title{Raman Sideband Thermometry of Single Carbyne Chains}

\author{Cla Duri Tschannen}
\email{clat@ethz.ch}
\author{Martin Frimmer}
\affiliation{Photonics Laboratory, ETH Zürich, 8093 Zürich, Switzerland}

\author{Georgy Gordeev}
\affiliation{Department of Physics, Freie Universität Berlin, 14195 Berlin, Germany}

\author{Thiago L. Vasconcelos}
\affiliation{Divisão de Metrologia de Materiais,
Instituto Nacional de Metrologia Qualidade e
Tecnologia (INMETRO), 25250-020 Duque de Caxias, RJ, Brazil}

\author{Lei Shi}
\affiliation{School of Materials Science and Engineering, State Key Laboratory of Optoelectronic Materials and Technologies, Nanotechnology Research Center, Sun Yat-sen University, Guangzhou 510275, Guangdong, P. R. China}

\author{Thomas Pichler}
\affiliation{Faculty of Physics, Universität Wien, 1090 Wien, Austria}

\author{Stephanie Reich}
\affiliation{Department of Physics, Freie Universität Berlin, 14195 Berlin, Germany}

\author{Sebastian Heeg}
\affiliation{Department of Physics, Freie Universität Berlin, 14195 Berlin, Germany}
\affiliation{Department of Physics, Humboldt Universität zu Berlin, 12489 Berlin, Germany}

\author{Lukas Novotny}
\affiliation{Photonics Laboratory, ETH Zürich, 8093 Zürich, Switzerland}

\begin{abstract}
We demonstrate Raman sideband thermometry of single carbyne chains confined in double-walled carbon nanotubes. Our results show that carbyne's record-high Raman scattering cross section enables anti-Stokes Raman measurements at the single chain level. Using laser irradiation as a heating source, we exploit the temperature dependence of the anti-Stokes/Stokes ratio for local temperature sensing. Due to its molecular size and its large Raman cross section carbyne is an efficient probe for local temperature monitoring, with applications ranging from nanoelectronics to biology.
\end{abstract}

\maketitle

\section{Introduction}\label{sec:level1}
Raman spectroscopy is a versatile non-destructive characterization method that is widely used in nanomaterials research owing to its sensitivity to structural, electronic, optical, and chemical information~\cite{Thomsen2007-hz, Jorio2011-zt, Saito2016-dr, Cong2020-cj}. At the heart of Raman spectroscopy lies the inelastic scattering of light by quantized lattice vibrations (phonons)~\cite{Raman1928-bi}. In the Raman process, an incoming photon can either create (Stokes) or annihilate (anti-Stokes) a phonon. The anti-Stokes process, being related to the annihilation of a phonon, can only occur if a phonon is already present in the system. The anti-Stokes scattering intensity is therefore proportional to the phonon population $n$, which is governed by the local temperature. Stokes Raman scattering, on the other hand, can also occur in the absence of a phonon. As a result, the Stokes Raman scattering intensity is proportional to $n + 1$ and thus the powers scattered into the anti-Stokes and the Stokes sideband differ. This sideband asymmetry provides direct access to the local temperature of the scattering material~\cite{Cardona1982-le, Jorio2011-zt}. However, for phonon energies larger than the thermal energy, phonon population numbers are very small ($n\ll1$). Consequently, anti-Stokes signals of high-energy ($>$ \SI{1000}{\per\centi\meter}) 
Raman modes are several orders of magnitude weaker than their Stokes counterparts. This discrepancy explains why for nanoscale systems, anti-Stokes spectra of high-energy Raman modes are only accessible for bulk quantities~\cite{Brown2000-di}, in solution~\cite{Gordeev2017-yu}, or under exotic conditions such as surface-enhanced Raman scattering~\cite{Maher2006-fu}, bias-induced hot phonon generation~\cite{Oron-Carl2008-ub}, or correlated Stokes--anti-Stokes scattering~\cite{Jorio2014-tz}. Raman sideband thermometry using a high-energy mode of an isolated system of molecular size has, to the best of our knowledge, not been demonstrated yet. 

A promising candidate to close this gap is carbyne, the paradigmatic sp-hybridized and truly one-dimensional allotrope of carbon~\cite{Heimann1999-ya, Hirsch2010-vr}. Carbyne has a resonant Raman scattering cross section per atom exceeding that of any other known material or molecule by at least two orders of magnitude~\cite{Tschannen2020-sy}. This record-high Raman cross section suggests that the temperature dependence of carbyne's C-mode optical phonon population can be exploited for local temperature measurements, despite the low occupation number associated with the high energy of the C-mode ($n\approx1.4\times10^{-4}$ at room temperature). Carbyne-based Raman sideband thermometry could therefore provide for a practical method for \textit{in situ} temperature monitoring on the nanoscale.

\begin{figure*}[t]
\includegraphics[width=\linewidth]{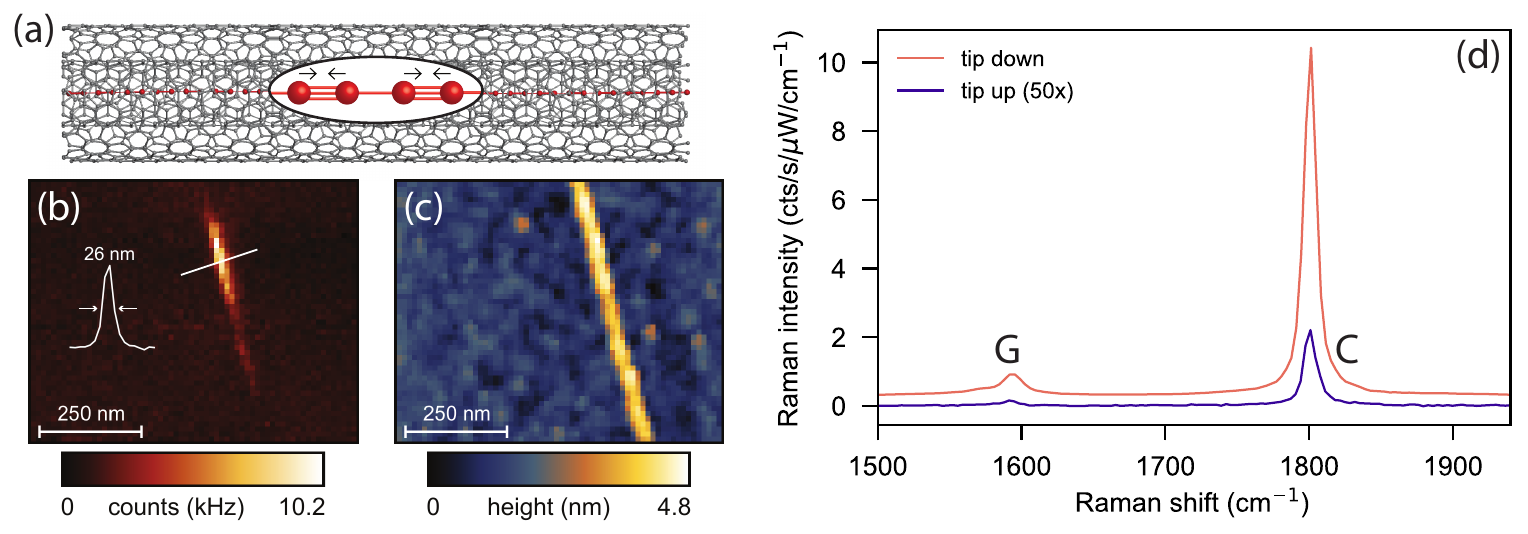}% Here is how to import EPS art
\caption{\label{fig:Figure1} Tip-enhanced Raman scattering of confined carbyne. (a)~Carbyne chain encapsulated in a DWCNT. The inset illustrates the atomic structure of carbyne, which is characterized by alternating single and triple bonds. Arrows illustrate the Raman-active vibrational C-mode. (b--c)~Simultaneously acquired near-field (b) and topographic (c) image of a single confined carbyne chain. The intensity profile in (b) extracted along the white line indicates a spatial resolution of \SI{26}{nm}. (d)~Near-field (tip down) and far-field (tip up) Raman spectrum of confined carbyne. The strong C-peak at \SI{1801}{\per\centi\m} is associated with the carbyne chain, whereas the G-peak signature below \SI{1600}{\per\centi\meter} arises from the encapsulating DWCNT structure. Note that the far-field spectrum is scaled by a factor of 50 for better visibility.}
\end{figure*}

Carbyne has attracted significant interest due to its anticipated outstanding mechanical~\cite{Liu2019-hk}, thermal~\cite{Wang2015-rl}, and electronic~\cite{Tongay2004-fk} properties. In particular, carbyne possesses a band gap that is sensitive to external perturbations~\cite{Shi2017-zk, Heeg2018-wp, Heeg2018-db, Yang2020-dg, Sharma2020-ej} and can possibly be switched externally from a semiconducting to a metallic state~\cite{PhysRevB.72.155420, PhysRevB.94.195422}, which points out the potential of carbyne for nanoelectronic devices. While the synthesis of carbyne has long been challenging~\cite{Gibtner2002-gp, Chalifoux2010-ea, Casari2018-mz}, large progress has been made in recent years using carbon nanotubes as nanoreactors and protective nanocontainers~\cite{Zhao2003-iz,Fantini2006-wu, Shi2011-sj, Andrade2015-ml, Shi2016-oj, Zhang2018-ek, Toma2019-tb, Shi2021-hj}, as illustrated in Figure 1a. Raman studies have played an important role in the investigation of carbyne, but have so far been restricted to Stokes scattering~\cite{Heeg2018-db, Heeg2018-wp, Shi2017-zk, Fantini2006-wu, Andrade2015-ml}. Extending this scope to the anti-Stokes side of the spectrum is not only of practical relevance for sideband thermometry, but can aid in assessing fundamental material properties such as phonon lifetimes~\cite{Song2008-bb} and optical transition energies~\cite{Souza_Filho2004-xx}, as well as extending the mechanistic understanding of the Raman process~\cite{Gordeev2017-yu}. 

In this Letter, we demonstrate that carbyne's unrivaled Raman scattering cross section enables anti-Stokes measurements at the single chain level. Using laser irradiation as a heating source, we investigate the temperature dependence of the anti-Stokes/Stokes ratio and demonstrate the applicability of carbyne as a nanoscale temperature sensor. 

\section{\label{sec:level2}Experimental}
Carbyne chains are synthesized inside double-walled carbon nanotubes (DWCNTs) by high-temperature annealing according to the procedure described by Shi \textit{et al.}~\cite{Shi2016-oj}. The tubes are then dispersed on a thin glass coverslip following Ref.~\citenum{Heeg2018-wp}.

\begin{figure*}[t]
\includegraphics[width=\linewidth]{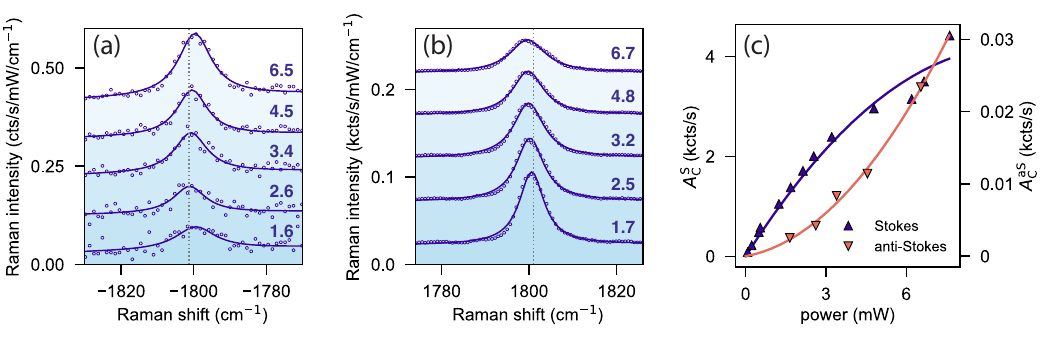}% Here is how to import EPS art
\caption{\label{fig:Figure2} Power-dependent anti-Stokes and Stokes Raman spectroscopy of an individual confined carbyne chain. (a)~anti-Stokes and (b)~Stokes spectra recorded at different excitation powers, which are indicated in blue in units of mW. The spectra are normalized by the excitation power and offset vertically for better visibility. The lines are Lorentzian fits on a linear background. The grey dotted lines at $\pm$\SI{1801}{\per\centi\meter} serve as reference to illustrate changes in the peak position. (c)~Stokes ($A_{\mathrm{C}}^{\mathrm{S}}$) and anti-Stokes ($A_{\mathrm{C}}^{\mathrm{aS}}$) Raman signals as a function of excitation power. The lines are second-order polynomial fits.}
\end{figure*}

Our work makes use of both near-field and far-field measurements, each carried out on a separate Raman setup. The first step is to identify an isolated, continuous carbyne chain confined in a DWCNT. To this end, we make use of the nanoscale resolution provided by tip-enhanced Raman scattering (TERS). The TERS measurements are performed in a back-scattering configuration using a home-built setup~\cite{Hartschuh2003-iu, Lapin2015-hv}. As optical probe we use a plasmon-tunable tip pyramid designed to support localized surface plasmon resonances at the excitation energy of our laser (1.96 eV), which gives rise to particularly strong Raman scattering enhancement~\cite{Vasconcelos2018-kt, Oliveira2020-ph, Miranda2020-dc}. Raman scattered photons are sent through a narrow band-pass filter that transmits only the spectral region of carbyne's C-mode and then detected with an avalanche photodiode (APD). Optical and topographic images are formed simultaneously by raster-scanning [see Figure~\ref{fig:Figure1}(b--c)]. Full spectra are recorded using a CCD-equipped spectrometer [Figure~\ref{fig:Figure1}(d)]. 

In the second experimental step, we perform power-dependent far-field Raman measurements of the previously identified carbyne chain. The measurements are carried out in a backscattering geometry with a 100x air objective (NA 0.9). We align both the polarization of the excitation laser and the direction of polarized detection with the chain's main axis, as detailed in Ref.~\cite{Heeg2018-wp}. A tunable dye laser serves as excitation source. To profit from resonant Raman scattering enhancement, we choose the photon energy of the laser to coincide with the band gap of the investigated carbyne chain (1.88 eV). We infer the band gap energy by exploiting its linear scaling with the measured C-mode Raman shift [1801 cm$^{-1}$ according to Figure~\ref{fig:Figure1}(d)]~\cite{Shi2017-zk, Heeg2018-wp, Heeg2018-db}. We point out that the band gap energy and Raman peak position of confined carbyne are both affected by the noncovalent interaction with the encasing host nanotube, which in turn depends on the nanotube chirality. Therefore, carbyne chains encapsulated in nanotubes of different chirality differ in their vibrational and electronic properties~\cite{Shi2017-zk, Heeg2018-wp, Heeg2018-db}.

For all power-dependent far-field measurements, we continuously monitor the power during signal acquisition. All power values reported in this work refer to mean values (averaged over the integration time) as measured before the back aperture of the objective. Also, instead of linearly varying the laser power, we alternate between higher and lower power values. This counteracts any systematic effects introduced by possible drifts of the sample with respect to the laser focus. The integration time for all Stokes (anti-Stokes) measurements is \SI{60}{s} (\SI{180}{s}).

\section{Results and Discussion}
We first analyze anti-Stokes and Stokes spectra of a single confined carbyne chain, then discuss the power dependence of the anti-Stokes and Stokes Raman signals, and finally quantify the laser-induced heating using Raman sideband thermometry. 

Figure~\ref{fig:Figure2} shows our power-dependent far-field Stokes and anti-Stokes measurements. A series of anti-Stokes spectra, measured at five different values of laser power, are displayed in Figure~\ref{fig:Figure2}(a). We fit each spectrum with a Lorentzian lineshape on top of a linear background. In Figure~\ref{fig:Figure2}(b) we display the corresponding series of Stokes spectra. We find equal Raman shifts for the Stokes and anti-Stokes peaks, as expected for a first-order single-resonance Raman process~\cite{Cardona1982-le, Jorio2011-zt}. Moreover, we observe a softening of the C-mode Raman shift with increasing laser power. Such frequency downshifts have been observed in temperature- and power-dependent Raman studies of a wide variety of materials~\cite{Calizo2007-lk, Sahoo2013-ut, Late2014-by, Liu2019-hk, Viana2020-ke} and can be attributed to anharmonic terms of the lattice potential energy~\cite{Jorio2011-zt}. We provide a brief discussion of this effect in the Supporting Information.

To shed light on the power dependence of the Raman signals, we extract the integrated C-peak anti-Stokes intensities $A_\mathrm{C}^\mathrm{aS}$ from the fits in Figure~\ref{fig:Figure2}(a), and the Stokes intensities $A_\mathrm{C}^\mathrm{S}$ from the fits in Figure~\ref{fig:Figure2}(b). We plot these Stokes and anti-Stokes signals in Figure~\ref{fig:Figure2}(c) as a function of excitation power. To illustrate the power-dependent behavior of the Stokes and anti-Stokes signals, we fit both sets of data with a second order polynomial, plotted in Figure~\ref{fig:Figure2}(c) as solid lines. Inspection of Figure~\ref{fig:Figure2}(c) demonstrates that the anti-Stokes and Stokes Raman signals show opposing trends with increasing excitation power. The anti-Stokes signal $A_{\mathrm{C}}^{\mathrm{aS}}$ increases supra-linearly, which can be ascribed to the temperature dependence of the C-mode phonon population given by the Bose-Einstein distribution, ${n= \{\exp{[E_{\mathrm{ph}}/(k_{\mathrm{B}}T)]}-1\}^{-1}}$, where $E_{\mathrm{ph}}$ is the phonon energy, $k_\mathrm{B}$ is the Boltzmann constant, and $T$ the temperature. Laser-induced heating increases the phonon occupation number $n$, which in turn gives rise to stronger anti-Stokes scattering. The Stokes signal $A_{\mathrm{C}}^{\mathrm{S}}$, on the other hand, grows sub-linearly with laser power. We attribute this behavior to changes of the underlying Raman resonance due to laser heating. These changes can include a shift of the optical transition energy as well as an enhanced non-radiative damping of the photoexcited state. Models including such effects have been frequently employed in power- and temperature-dependent Raman studies of low-dimensional semiconductors~\cite{Livneh2010-wq, Fan2014-nw, Zobeiri2020-ew}. Moreover, this explanation is in line with the strong C-mode Raman intensity increase observed by Shi \textit{et al.} at cryogenic temperatures in bulk measurements~\cite{Shi2016-oj}. We emphasise that while the exact mechanism behind the saturation of the Stokes Raman signal towards high excitation powers in Figure~\ref{fig:Figure2}(c) requires further investigation, it has no relevant implications for Raman sideband thermometry. This robustness of the anti-Stokes/Stokes ratio of carbyne against changes of the electronic structure is a consequence of the large C-mode phonon energy, as elaborated in the final part of this section.

\begin{figure}[t]
\includegraphics[width=0.9\linewidth]{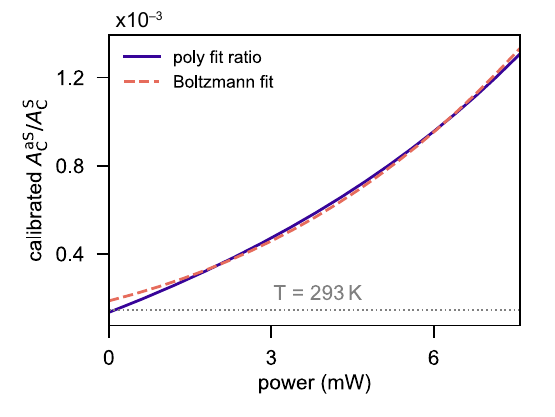}% Here is how to import EPS art
\caption{\label{fig:Figure3} Anti-Stokes/Stokes ratio ($A_{\mathrm{C}}^{\mathrm{aS}}/A_{\mathrm{C}}^{\mathrm{S}}$) of a single confined carbyne chain as a function of excitation power. The blue line is the ratio of the two polynomial fits shown in Figure~\ref{fig:Figure2}(c). The red dashed line is a fit to the blue curve of the Boltzmann factor with a power-dependent equilibrium temperature, as explained in the main text. The grey dotted line represents the Boltzmann factor at room temperature (\SI{293}{K}).}
\end{figure}

After having discussed the power-dependence of the anti-Stokes and Stokes signals separately, we now focus on their ratio $A_{\mathrm{C}}^{\mathrm{aS}}/A_{\mathrm{C}}^{\mathrm{S}}$ and its power dependence. In Figure~\ref{fig:Figure3} we plot the ratio of the two polynomial fits from Figure~\ref{fig:Figure2}(c) as a solid line. This curve represents our experimental anti-Stokes/Stokes ratio as a function of laser power. Note that going from Figure~\ref{fig:Figure2}(c) to Figure~\ref{fig:Figure3}, we have applied corrections of the anti-Stokes/Stokes ratio for the $\omega^4$~dependence of Raman scattering~\cite{Cardona1982-le, Jorio2011-zt, Novotny2012-se} as well as for the frequency dependence of the instrument. 

With this calibration, we expect the anti-Stokes/Stokes ratio to match the Boltzmann factor $A_{\mathrm{C}}^{\mathrm{aS}}/A_{\mathrm{C}}^{\mathrm{S}} = \exp{[-E_{\mathrm{ph}}/(k_{\mathrm{B}}T)]}$~\cite{Cardona1982-le, Jorio2011-zt}. Indeed, as shown in Figure~\ref{fig:Figure3}, we find that laser irradiation increases the anti-Stokes/Stokes ratio and the associated phonon equilibrium temperature $T$. Assuming that the temperature scales linearly with excitation power $P_{\mathrm{L}}$, we can extract the rate $C_{\mathrm{T}}$ of laser-induced heating from fitting a modified Boltzmann factor, $B\exp{\{-E_{\mathrm{ph}}/[k_{\mathrm{B}}(293 + C_{\mathrm{T}}P_{\mathrm{L}})]\}}$. The fit is shown as the red dashed line in Figure~\ref{fig:Figure3} and yields a heating rate of $C_{\mathrm{T}} = 11.1$~($+$2.4/$-$1.8)~\si{K/mW}. The free parameter $B$, for which we obtain a value of 1.29, accounts for uncertainties of the calibration procedure. When extrapolated to zero power, both the anti-Stokes/Stokes ratio and the Boltzmann fit in Figure~\ref{fig:Figure3} closely coincide with the room-temperature Boltzmann factor (grey dotted line). This finding supports the accuracy of our calibration and corroborates that we excite the carbyne chain near its electronic resonance~\cite{Heeg2018-wp, Brown2000-di, Fantini2004-iz}. Moreover, from the agreement of the two curves in Figure~\ref{fig:Figure3}, we conclude that heating is the main contribution to the anti-Stokes/Stokes ratio as the laser power is increased. Thus, correlated Stokes--anti-Stokes scattering (a nonlinear process) does not play a sizeable role~\cite{Jorio2014-tz, Parra-Murillo2016-rz}. Based on our estimate of the heating rate $C_{\mathrm{T}}$, the power range depicted in Figure~\ref{fig:Figure3} corresponds to a temperature increase of roughly \SI{90}{\kelvin}, which leads to the observed increase of the anti-Stokes/Stokes ratio by one order of magnitude. 

In the Supporting Information we provide two separate consistency checks for our results. First, we correlate the laser-heating induced C-mode shifts with temperature-dependent measurements, which allows us to validate the heating rate estimate derived from the power-dependent anti-Stokes and Stokes measurements. Second, we reproduce Figures~\ref{fig:Figure2}(c) and~\ref{fig:Figure3} with another carbyne chain of similar C-mode frequency and find excellent agreement between the two carbyne chains.

Let us discuss the prospects of single carbyne chains for nanoscale temperature sensing. In our work, we use resonant Raman scattering, that is, our excitation frequency is resonant with the carbyne band gap. As discussed above, we argue that the temperature dependence of this resonance with respect to energetic position and width explains the sub-linear dependence of the Stokes Raman signal on excitation power [see Figure~\ref{fig:Figure2}(c)]. Importantly, a change in resonance influences the Stokes and anti-Stokes signals of carbyne to the same extent, thus rendering their ratio immune to changes of the resonance's position and width. This robustness persists as long as the resonance shift remains small compared to the phonon energy. In this regime, only the resonance condition for the excitation laser plays a role (incoming resonance)~\cite{Heeg2018-wp}. However, too large of a temperature-induced resonance shift will bring the Stokes or anti-Stokes emission into resonance (outgoing resonance). In that regime, the sideband ratio starts to deviate from the Boltzmann factor~\cite{Brown2000-di,Fantini2004-iz}, which complicates the extraction of a temperature. 

From our measurements, we extract a heating-induced shift of the C-mode Raman peak of around \SI{-2}{\per\centi\meter} [see Figure~\ref{fig:Figure2}(b) and Supporting Information], corresponding to a thermal shift of the band gap energy of \SI{-13}{meV}~\cite{Shi2017-zk, Heeg2018-db}. We underline that the direct correlation between Raman frequency and optical transition energy is an inherent property of confined carbyne~\cite{Shi2017-zk, Heeg2018-wp, Casari2018-mz} that sets carbyne apart from other nanomaterials for which the influence of temperature on electronic structure is more difficult to assess~\cite{Zobeiri2020-ew, Fan2014-nw, Livneh2010-wq, Zhou2006-gx, Zhang2007-ph}. Given the much larger C-mode phonon energy of more than \SI{220}{meV}, the measured temperature-induced downshift of the band gap by \SI{13}{meV} has no relevant effect on the anti-Stokes/Stokes ratio of carbyne. Hence, in the context of Raman sideband thermometry, large phonon energies allow for reliable and robust temperature extraction from the anti-Stokes/Stokes ratio, regardless of changes of the electronic structure due to temperature variations. While anti-Stokes signals of high-energy phonons are in general challening to obtain due to low phonon occupation numbers, carbyne overcomes this difficulty by virtue of its record-high Raman scattering cross-section~\cite{Tschannen2020-sy}, allowing for anti-Stokes measurements---and thus, robust sideband thermometry---at the single chain level.

\section{Conclusion and Outlook}
In conclusion, we have experimentally shown that the unparalleled Raman scattering cross section of carbyne~\cite{Tschannen2020-sy} enables anti-Stokes spectroscopy at the single chain level. Using laser irradiation as a heating source, we have investigated the temperature dependence of the Stokes and anti-Stokes Raman peaks. Finally, we have demonstrated that the anti-Stokes/Stokes ratio of carbyne provides a means for nanoscale temperature sensing. Our work therefore establishes a practical method for all-optical probing of temperature variations on the nanoscale, with potential applications in many areas of modern science and technology. 

For instance, carbyne-based Raman sideband thermometry offers a promising perspective for thermal management on the length scale of emerging ultracompact device architectures, which is a key requirement for their functioning and control~\cite{Balandin2002-kw, Cahill2003-iw, Van_Erp2020-yg}. This prospect bears particular relevance in light of the recent progress in the synthesis of carbyne chains with tailored properties, which puts the integration of carbyne into the next generation of nanoelectronic devices firmly within reach~\cite{Shi2018-fu, Chimborazo2019-xy, Shi2021-hj}.

The integration of carbyne within a living system offers further intriguing possibilities, ranging from tuning gene expression by simultaneously monitoring and controlling subcellular thermal gradients~\cite{Kamei2009-yz, Xu2012-lo, Kumar2010-hd} to investigating local tumor activity by mapping atypical thermogenesis at the single-cell level~\cite{Lemos2019-nq}. On top, the demonstrated robustness of carbyne towards large-power continuous-wave laser irradiation for prolonged integration times adds to the list of advantages that would make it a powerful marker for use in biological imaging~\cite{Tschannen2020-sy}. 

Finally, our measurements constitute a starting point for the further investigation of anti-Stokes scattering in carbyne. In particular, exploring the outgoing anti-Stokes resonance might not only increase the absolute anti-Stokes signal, as shown for carbon nanotubes~\cite{Gordeev2017-yu}, but also shed more light on the exact nature of optical transitions in carbyne~\cite{Thomsen2007-hz,Tran2016-ov, Heeg2018-db}.

\begin{acknowledgements}
We thank Ben Weintrub for help with the temperature stage. This work has been supported by the Swiss National Science Foundation (Grant 200020\_192362/1). T.L.V acknowledges financial support from CNPq (305881/2019-1 and 436381/2018-4) and MCTI (SibratecNano 21040*16). L.S. acknowledges the financial support from the National Natural Science Foundation of China (Grant 51902353) and Natural Science Foundation of Guangdong Province (Grant 2019A1515011227). S.H. gratefully acknowledges funding from the Deutsche Forschungsgemeinschaft through the Emmy Noether Programme (HE 8642/1-1).
\end{acknowledgements}


\bibliography{apssamp.bib}

\end{document}